\documentclass[conference]{IEEEtran}
\IEEEoverridecommandlockouts

\IEEEpubid{\makebox[\columnwidth]{000-0-0000-0000-0/00/\$00.00~\copyright2026 IEEE \hfill} \hspace{\columnsep}\makebox[\columnwidth]{ }}

\usepackage{cite}
\usepackage{amsmath,amssymb,amsfonts}
\usepackage{algorithmic}
\usepackage{graphicx}
\usepackage{textcomp}
\usepackage{xcolor}
\usepackage{url}
\def\BibTeX{{\rm B\kern-.05em{\sc i\kern-.025em b}\kern-.08em
    T\kern-.1667em\lower.7ex\hbox{E}\kern-.125emX}}
\begin{document}

\title{Turning Language Model Training \\
from Black Box into a Sandbox
\thanks{This study was funded by the Strategic Research Council (SRC) established within the Research Council of Finland, grant \#352859.}
}

\author{
\IEEEauthorblockN{Nicolas Pope}
\IEEEauthorblockA{\textit{School of Computing} \\
\textit{University of Eastern Finland}\\
Joensuu, Finland \\
npope@uef.fi}
\and
\IEEEauthorblockN{Matti Tedre}
\IEEEauthorblockA{\textit{School of Computing} \\
\textit{University of Eastern Finland}\\
Joensuu, Finland \\
matti.tedre@uef.fi}
}

\maketitle

\IEEEpubidadjcol

\begin{abstract}

Most classroom engagements with generative AI focus on prompting pre-trained models, leaving the role of training data and model mechanics opaque. We developed a browser-based tool that allows students to train a small transformer language model entirely on their own device, making the training process visible. In a CS1 course, 162 students completed pre- and post-test explanations of why language models sometimes produce incorrect or strange output. After a brief hands-on training activity, students' explanations shifted significantly from anthropomorphic and misconceived accounts toward data- and model-based reasoning. The results suggest that enabling learners to directly observe training can support conceptual understanding of the data-driven nature of language models and model training, even within a short intervention.  For K-12 AI literacy and AI education research, the study findings suggest that enabling students to train---and not only prompt---language models can shift how they think about AI.
\end{abstract}


\section{Introduction}

Large language models (LLMs) have become everyday tools in professional, creative, and educational contexts. Yet for most learners, these systems remain opaque black boxes: tools to use rather than mechanisms to understand \cite{gu25,grover24}. Classroom activities often focus on prompting, evaluating, or auditing the behavior of industry scale models, leaving the processes that make them work hidden. Students may see polished generative outputs, but not the messy, iterative process of data curation and model training, and why models sometimes produce responses that are implausible, incoherent, or ``hallucinated.''  Critical AI literacy research argues that learners need opportunities to \textit{experience} how AI systems are constructed and shaped by human choices, rather than encountering them only as finished products \cite{kafai25,vartiainen25f,stetsenko19,veldhuis25,iivari24b}.  Making the mechanisms of LLM training visible supports conceptual understanding of their statistical, data-driven nature.

Making the training process visible, however, has proven difficult in the case of language models. Training LLMs requires specialized hardware, large-scale datasets, and cloud-based inference pipelines that are not available in schools and introductory university courses. Although browser-based educational tools such as Teachable Machine have successfully supported learning about image classification workflows \cite{carney20,pope24b}, similar pedagogically oriented tools for generative language models are scarce, especially tools that reveal the complete workflow from data selection to training and inference, are aimed at education, and use real transformer technology.

The \emph{Little Language Machine} addresses this gap. It is a browser-native learning environment in which students can train a miniature transformer-based model directly on their own devices. Learners upload or compose small text corpora, select model size, and observe training unfold in real time. Within minutes, the model progresses from random character noise to increasingly coherent text sequences, making visible how training data and duration shape the model's behavior. Because the tool runs entirely client-side using WebGPU acceleration, it ensures GDPR-compliance, no setup overhead, and works even on low-end classroom laptops.

\newcommand{\circlenum}[1]{\raisebox{.5pt}{\textcircled{\raisebox{-.9pt} {#1}}}} 
\newcommand{\smcirclenum}[1]{\raisebox{.5pt}{\textcircled{\raisebox{-.4pt} {\scriptsize{#1}}}}} 
\begin{figure*}[htbp]
\centering{\includegraphics[width=\textwidth]{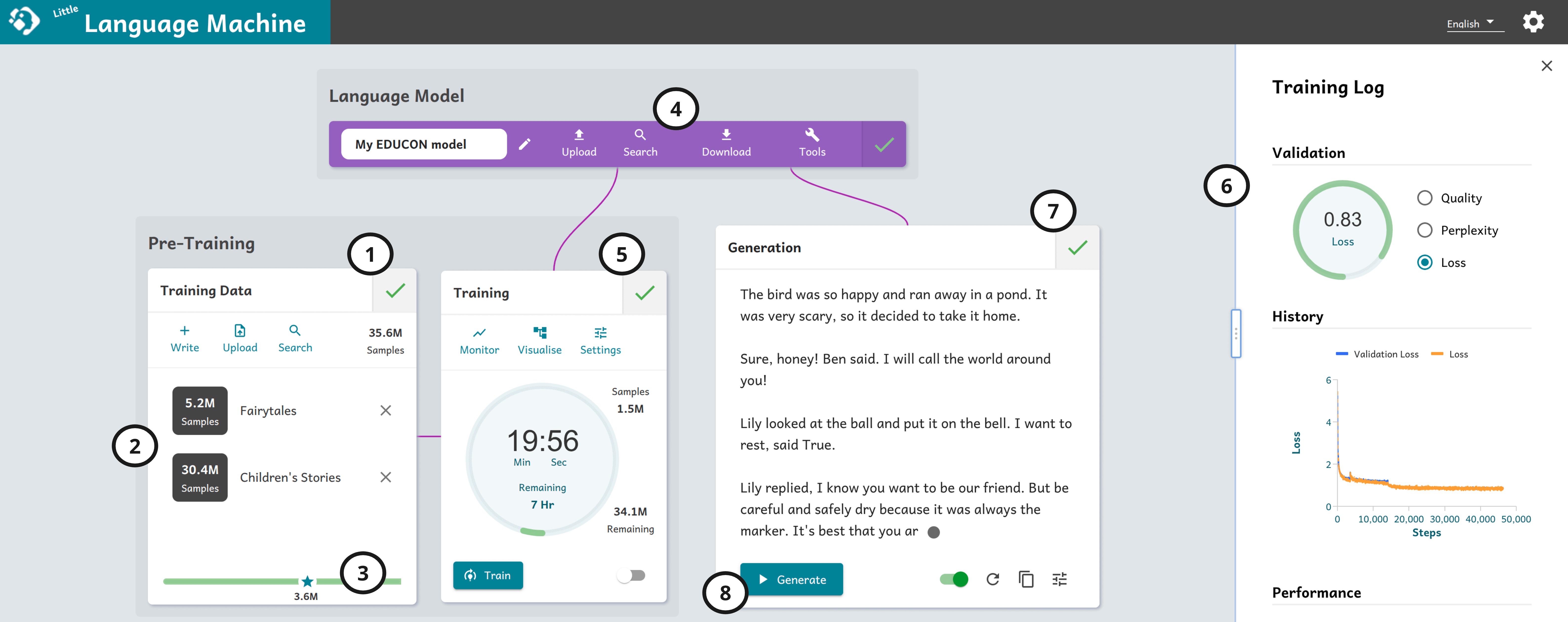}}
\caption{Screen shot of Little Language Machine.  When no XAI options are selected, the basic interface consists of five main windows: \smcirclenum{1} Training data selection, \smcirclenum{4} Model selection, \smcirclenum{5} Training process tracker, \smcirclenum{6} Model quality evaluation, and \smcirclenum{7} Text generator window.}
\label{fig:LLM}
\end{figure*}

This study deployed the Little Language Machine at scale in an Introduction to Computing (CS1) course. A total of 198 learners completed a pre-test and training activity, 222 did a post-test that measured changes in how they explained why language models sometimes produce incorrect or strange output.  Altogether 162 students completed both activities.  We analyzed students' evolving explanatory models to understand how this short, hands-on training experience shifted learners' reasoning about the mechanisms of language models.

\section{System Description}

The \textit{Little Language Machine} (\textcolor{blue}{\url{lm.gen-ai.fi}}) runs a modified nanoGPT-style transformer \cite{karpathy23} entirely within the browser.  The system supports different model sizes; for instance, one configuration consists of three self-attention heads, eight transformer blocks, 192 hidden units, and a 128-token context window, totaling 4 million parameters.  A tiny two-million-parameter, character-level model that occupies just 0.24GB memory is also available, designed to run within the memory constraints of typical student laptops and Chromebooks.  

The implementation uses root mean square layer normalization and rotary position embeddings (RoPE) for efficient positional encoding.  Training and inference are hardware-accelerated through WebGPU or WebGL, enabling fast training and inference on low-end devices. Key–value caching is used to maintain responsiveness during inference and mixed-precision training is used to reduce memory demands.  As no data are transmitted to external servers, the system is privacy-preserving, offline-capable, and GDPR-compliant by design.

Figure \ref{fig:LLM} illustrates the interaction workflow.  Learners begin by selecting or uploading a data set \smcirclenum{1}.  Ready-made corpora include, for example, children's short stories and Shakespeare's works \smcirclenum{2}.  A visual indicator \smcirclenum{3} displays whether the data set is sufficiently large for the chosen model.  The learner then selects a transformer configuration \smcirclenum{4}---either untrained, to watch learning unfold from noise to structure, or pre-trained for quick fine-tuning.  Training \smcirclenum{5} occurs directly in the browser, with loss and sample counts displayed in real time. Coherent text typically emerges within minutes.  

The system provides an automatic evaluation \smcirclenum{6} that encourages reflection on how data size, domain, and training time affect model behavior.  Learners can prompt the model at any stage in the generator window \smcirclenum{7} and generate new text \smcirclenum{8}.  Models can be exported or reloaded for further exploration or classroom sharing.  This end-to-end, in-browser workflow from data curation through training, evaluation, and generation, makes the inner workflow of language model training visible and manipulable, turning generative AI from a black box into a hands-on medium for inquiry and critique.

\section{Methods}

The study took place in a mandatory first-year CS1 (Introduction to Computing) course at a Finnish university.  The intervention consisted of a one-week sequence within the course. Students completed a pre-test consisting of two short open-ended questions asking how and why language models sometimes produce text that is incorrect, implausible, or nonsensical. Students were then assigned a structured homework task using the \emph{Little Language Machine}.  The task instructed learners to choose datasets, select model configurations, train and fine-tune models in the browser, observe how the output changed over time, and experiment with generating new text. Students submitted short written reflections describing what they observed about how and why the model's behavior changed during training.  Finally, students completed a post-test containing the same open-ended questions as the pre-test.  Importantly, the course curriculum had not yet introduced machine learning or language models at this point, meaning students' explanations were not shaped by prior formal instruction---only their assigned activities with the tool.

Participation in the learning activity was part of normal course work, but informed consent for inclusion in the research was requested separately, following national research ethics guidelines \cite{tenk23}. A total of 162 students completed the pre-test, homework activity, and post-test, as well as provided consent.  The primary data consisted of students' open-ended responses from the pre/post-tests. Responses were analyzed using qualitative content analysis \cite{elo08}. We developed an inductive coding scheme, iteratively coded the responses, and grouped them into eight conceptual categories representing different ways students explained why language models sometimes generate strange or incorrect text.  The goal of analysis was not to assess individual learning gains, but to characterize the explanatory models students used, and to understand how hands-on interaction with the system influenced those explanatory models. 

\section{Results and Discussion}

The 162 pre-test/post-test pairs were coded into eight conceptual categories representing how students explained why language models sometimes produce incorrect, implausible, or strange text:

\begin{enumerate}
    \item \textit{Training data}: Students attributed problems to deficiencies in the training data (e.g., quantity, quality, biases).
    \item \textit{Model deficiencies}: Students attributed problems to how the model works (e.g. their stochastic, predictive nature).
    \item \textit{Bad prompt}: Students attributed the problem to prompt.
    \item \textit{Bad search query}: Students believed that the LLM fails to search for information on the web.
    \item \textit{Translation issues (Eng/Fin)}: Students believed that the LLM fails to translate English results to Finnish.
    \item \textit{Lack of intelligence}: Students attributed the problems to the system's lack of human type intelligence. 
    \item \textit{Hallucinations}: Students blamed hallucinations.
\end{enumerate}

\newcommand{\graphwidth}[0]{0.7}

\begin{figure}[htbp]
\centering{\includegraphics[width=\graphwidth\columnwidth]{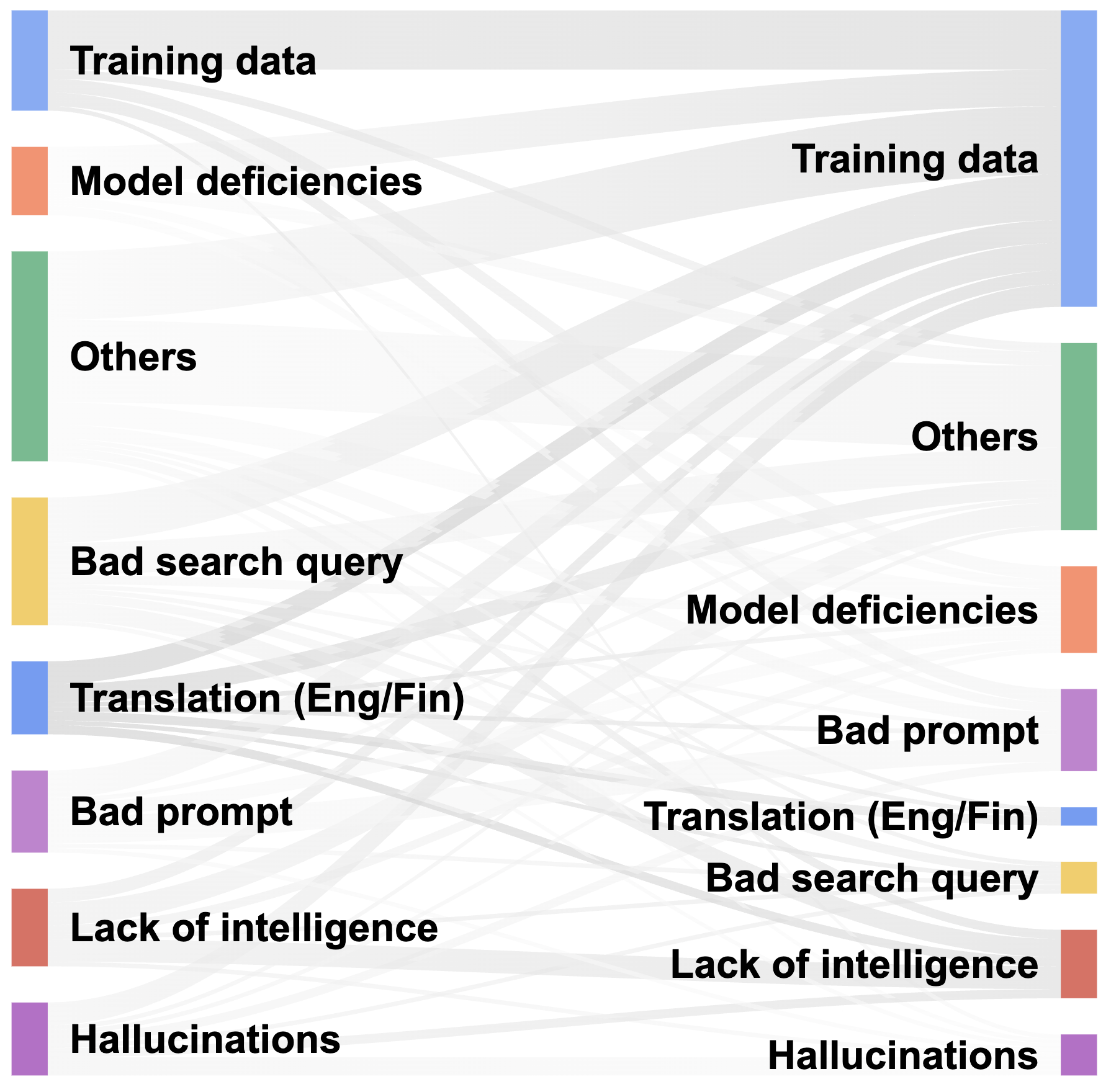}}
\caption{Students' responses to the task on  how and why language models sometimes produce text that is incorrect, implausible, or nonsensical.}
\label{fig:LLM_mistakes}
\end{figure}

Figure \ref{fig:LLM_mistakes} visualizes category transitions from pre- to post-test.  A clear shift occurred from explanations that described the model as an autonomous or intelligent agent toward explanations that emphasized data and the training process.  Misconceptions, such as believing that the model ``searches the web'' or ``mistranslates from English to Finnish'' decreased substantially (from 37 to 10 mentions), suggesting that the intervention influenced conceptual models of \textit{how} language models generate text.  Meanwhile, attributions to training data, such as training time, dataset size, quality, and bias, rose from 20 to 62 mentions.  Students used much more discipline-specific language, frequently describing that ``the training data set is too small or training data too short.''  Some students continued to refer to prompting issues, model stochasticity, or hallucinations, yet these were often described in ways more aligned with how language models function.  Quantitatively, the proportion of students attributing incorrect model output to training data increased from 13\% in the pre-test to 38\% in the post-test. A two-proportion z-test confirmed this increase to be statistically significant ($z = 5.09, p < .001$), indicating a meaningful conceptual shift toward data-centered reasoning.


Taken together, the results indicate that even brief, hands-on experience training a small transformer model can help students move away from common misconceptions and anthropomorphic explanations toward  disciplinary accounts of model behavior. While the intervention alone falls short of producing complete or fully expert mental models, it makes the statistical and data-dependent nature of language models visible and discussable, supporting recent calls for AI literacy approaches that treat learners as investigators and builders of AI systems rather than passive users \cite{vartiainen25f,kafai25}.  This supports the value of integrating transparent, trainable AI systems into introductory computing education.

\section{Conclusion}
This pilot study shows that enabling students to train a language model can shift their explanations of model behavior toward data-centered reasoning. By making the training process visible and manipulable, the \emph{Little Language Machine} helps learners understand language models as statistical systems shaped by training data. Integrating such transparent, hands-on systems into introductory computing courses can support more grounded and critical forms of AI literacy, shifting from ``prompting'' black-box systems to hands-on model training.

\bibliographystyle{IEEEtran}
\bibliography{bibliography}

@inproceedings{gu25,
	Address = {New York, NY, USA},
	Author = {Gu, Xingjian and Ericson, Barbara J.},
	Booktitle = {Proceedings of the 2025 ACM Conference on International Computing Education Research V.1},
	Doi = {10.1145/3702652.3744217},
	Isbn = {9798400713408},
	Numpages = {16},
	Pages = {125--140},
	Publisher = {ACM},
	Series = {ICER '25},
	Title = {{AI} Literacy in {K-12} and Higher Education in the Wake of Generative {AI}: An Integrative Review},
	Url = {https://doi.org/10.1145/3702652.3744217},
	Year = {2025}}

@webpage{karpathy23,
	Author = {Karpathy, Andrej},
	Title = {{nanoGPT}},
	Url = {https://github.com/karpathy/nanoGPT},
	Year = {2023}}

@inproceedings{vartiainen25f,
	Address = {Joensuu, Finland},
	Author = {Vartiainen, Henriikka and Tedre, Matti},
	Booktitle = {Proceedings of the International Conference on Smart Learning Environments},
	Month = {October 16--17},
	Title = {Towards Transformative {AI} Education Through Cross-Boundary Co-Design},
	Year = {2025}}

@article{veldhuis25,
	Author = {Annemiek Veldhuis and Priscilla Y. Lo and Sadhbh Kenny and Alissa N. Antle},
	Doi = {https://doi.org/10.1016/j.ijcci.2024.100708},
	Issn = {2212-8689},
	Journal = {International Journal of Child-Computer Interaction},
	Pages = {100708},
	Title = {Critical Artificial Intelligence literacy: A scoping review and framework synthesis},
	Url = {https://www.sciencedirect.com/science/article/pii/S2212868924000771},
	Volume = {43},
	Year = {2025}}

@inproceedings{iivari24b,
	Address = {New York, NY, USA},
	Author = {Iivari, Netta and Iversen, Ole Sejer and Smith, Rachel Charlotte and Schaper, Marie-Monique and Vent\"{a}-Olkkonen, Leena and Hartikainen, Heidi and Sharma, Sumita and Kinnula, Marianne and Lehto, Essi and Holappa, Jenni and Molin-Juustila, Tonja},
	Booktitle = {Proceedings of the 23rd Annual ACM Interaction Design and Children Conference},
	Doi = {10.1145/3628516.3655806},
	Isbn = {9798400704420},
	Location = {Delft, Netherlands},
	Numpages = {16},
	Pages = {322--337},
	Publisher = {ACM},
	Series = {IDC '24},
	Title = {Transformative agency -- the next step towards children's computational empowerment},
	Url = {https://doi.org/10.1145/3628516.3655806},
	Year = {2024}}

@article{stetsenko19,
	Author = {Stetsenko, Anna},
	Doi = {10.3389/feduc.2019.00148},
	Issn = {2504-284X},
	Journal = {Frontiers in Education},
	Pages = {1--13},
	Title = {Radical-Transformative Agency: Continuities and Contrasts With Relational Agency and Implications for Education},
	Volume = {4},
	Year = {2019}}

@inproceedings{kafai25,
	Author = {Kafai, Yasmin and Shapiro, R. Benjamin and Jetzinger, Franz and Michaeli, Tilman and Tedre, Matti and Vartiainen, Henriikka and Iivari, Netta and Musaeus, Line Have and Iversen, Ole Sejer and Ali, Safinah and Bodon, Herminio and Butler, Meg and Kshirsagar, Khushbu and Smith, Michael and Quiterio, Ashley and Worsley, Marcelo and Kumar, Vishesh and Morales-Navarro, Luis and Noh, Daniel and Pea, Roy and Philip, Thomas M.},
	Booktitle = {Proceedings of the 19th International Conference of the Learning Sciences},
	Organization = {ICLS},
	Pages = {2260-2268},
	Series = {ICLS 2025},
	Title = {Youth as Designers of Artificial Intelligence and Machine Learning Technologies: What Do We Know About the Opportunities and Challenges of {K-12} Students Creating Their Own Applications?},
	Year = {2025}}

@inproceedings{carney20,
	Address = {New York, NY, USA},
	Author = {Carney, Michelle and Webster, Barron and Alvarado, Irene and Phillips, Kyle and Howell, Noura and Griffith, Jordan and Jongejan, Jonas and Pitaru, Amit and Chen, Alexander},
	Booktitle = {The 2020 CHI Conference on Human Factors in Computing Systems},
	Doi = {10.1145/3334480.3382839},
	Location = {Honolulu, HI, USA},
	Numpages = {8},
	Pages = {1--8},
	Publisher = {ACM},
	Series = {CHI EA '20},
	Title = {Teachable Machine: Approachable Web-Based Tool for Exploring Machine Learning Classification},
	Url = {https://doi.org/10.1145/3334480.3382839},
	Year = {2020}}

@misc{pope24b,
	Archiveprefix = {TechRxiv},
	Author = {Nicolas Pope and Juho Kahila and Henriikka Vartiainen and Matti Tedre},
	Doi = {10.36227/techrxiv.24320794.v1},
	Title = {Children's {AI} Design Platform for Making and Deploying {ML}-Driven Apps},
	Year = {2024}}

@book{tenk23,
	Author = {TENK},
	Number = {4/2023},
	Publisher = {{TENK}},
	Series = {Publications of the Finnish National Board on Research Integrity},
	Title = {The Finnish Code of Conduct for Research Integrity and Procedures for Handling Alleged Violations of Research Integrity in Finland},
}

@conference{grover24,
	Address = {Portland, OR, USA},
	Author = {Grover, Shuchi},
	Booktitle = {Proceedings of {ACM} Computer Science Education ({SIGCSE}) 2024 Conference},
	Organization = {ACM},
	Pages = {1--7},
	Title = {Teaching {AI} to {K}-12 Learners: Lessons, Issues, and Guidance},
	Year = {2024}}

@article{elo08,
	Author = {Elo, Satu and Kyng{\"a}s, Helvi},
	Doi = {10.1111/j.1365-2648.2007.04569.x},
	Journal = {Journal of advanced nursing},
	Number = {1},
	Pages = {107--115},
	Title = {The qualitative content analysis process},
	Volume = {62},
	Year = {2008}}
\end{document}